\documentstyle [12pt]{article}
\begin{document}
\author{{\it Leonid B. Litinsky}}
\title{{\bf Energy functional and fixed points\\ of a neural network}}
\date{{\small Russia, 142092, Troitsk, Moscow Region,\\
Institute for High Pressure Physics Russian Academy of Sciences\\
fax: /095/-334-0012, e-mail: litin@ns.hppi.troitsk.ru}}
\maketitle
\begin{abstract}
A dynamic system, which is used in the neural network theory, Ising spin 
glasses and factor analysis, has been investigated. The properties of the 
connection matrix, which guarantee the coincidence of the set of the fixed 
points of the dynamic system with the set of local minima of the energy 
functional, have been determined. The influence of the connection matrix 
diagonal elements on the structure of the fixed points set has been 
investigated.
\end{abstract}
\section{Introduction}

We define a neural network as a dynamic system of $n$ spin variables 
({\sl spins}) which can take one of two values:
$$\sigma_i =\{\pm 1\},\quad i=1,2,\ldots,n. \eqno(1)$$
The spins are connected by a symmetric connection matrix $J=(J_{ij})$ :
$$J_{ij}=J_{ji},\quad i,j=1,2,\ldots,n.$$
The local potential
$$h_i(t)=\sum_{j=1}^n J_{ij}\sigma_j(t) \eqno(2)$$
with which the network acts on spin $i$, determines solely the value of spin 
$i$ at time $t+1$:

$$ \sigma_i(t+1)=\left\{\begin{array}{rcl} 
\sigma_i(t),&\mbox{ if }&h_i(t)\sigma_i(t)\ge 0\\ 
-\sigma_i(t),&\mbox{ if }&h_i(t)\sigma_i(t) < 0 \end{array}\right. \eqno(3)$$

The state of the network as a whole is described by a {\sl configuration vector}
$\vec \sigma$, whose coordinates are given by Eq.(1). In what follows, 
Greek letters will be used to designate configuration vectors.

We want to investigate the set of the so called {\sl fixed points} of the 
network, i.e. such states $\vec \sigma^*$, that for all coordinates $\sigma_i^*$ 
have:
$$\sigma_i^*h_i \ge 0,\quad i=1,2,\ldots,n. \eqno(4)$$

Besides the neural network theory, the mathematical model (1)-(3) is also used 
in the factor alalysis~\cite{Lit1} and in the Ising spin glass theory~\cite{Dom}. 
The neural network theory makes use of a physical concept of the 
{\sl energy of the state} $\vec \sigma$, which is defined as

$$E(\vec \sigma)=-\frac 1n\sum_{i=1}^n\sigma_ih_i=-\frac 1n \sum_{i,j=1}^n
J_{ij}\sigma_i\sigma_j. \eqno(5)$$                                

It is very important, both from the physical point of view and the ability of 
the network to have content-addressable memory~\cite{Dom}, that the energy of the 
state would be a decreasing function on every step of the network evolution. 
And, moreover, the fixed points must be the local minima of the energy 
functional (5). 

In the second section, we obtain the conditions under which a connection 
matrix $J$ guarantees the fulfillment of the above mentioned requirements. 
It has been found that the Hebb connection matrix as well as the connection 
matrices which are used in physical problems possess the necessary property. 
But this is not the case for the projection matrix~\cite{Pers}. As a result, a network 
with such a connection matrix has a set of fixed points which is wider than 
the set of the local minima of the energy functional (5). In the third 
section, we show how the situation for a network with a projection matrix 
can be improved.

{\bf Notations.}  In what follows, a network with a connection matrix $J$ is 
called a $J$-network. We denote by $FP(J)$ the set of all fixed points of the 
$J$-network. A configuration vector which is a fixed point of a network will 
have a superscript "*":
$$\vec \sigma^* \in FP(J).$$

We denote by $LM(J)$ the set of the local minima of the energy functional (5). 
To examine the local minima of the functional (5), we introduce a topology on 
the set of configuration vectors: the set of $n$ configuration vectors 
$\vec \sigma^{(l)}$ which are the nearest to the vector $\vec \sigma$ in the 
sense of the Hamming distance will be called a {\sl vicinity} of the state
$\vec \sigma$:
$$\vec \sigma^{(l)}=(\sigma_1,\sigma_2,\ldots,-\sigma_l,\ldots,\sigma_n), \quad
l=1,2,\ldots,n.$$
In other words, the state $\vec \sigma^{(l)}$ from the vicinity of the state 
$\vec \sigma$ differs from the latter by the opposite value of the $l$th 
spin only, 
$$\vec \sigma \in LM(J) \Leftrightarrow E(\vec \sigma) \le E(\vec \sigma^{(l)})
,\quad  l=1,2,\ldots,n.$$

Finally, a matrix with zero diagonal elements will be marked by the 
superscript "0":
$$J^{(0)}\Leftrightarrow J_{ii} =0,\quad i=1,2,\ldots,n.$$
\section{On the role of the diagonal elements}
        
{\bf Theorem 1}

{\bf 1.} The set of the local minima of the energy functional (5) does not depend 
on the value of the diagonal elements of the connection matrix:
$$LM(J)=LM(J+A),$$
where
$$A=diag\{a_{11},a_{22},\ldots,a_{nn}\} \eqno(6)$$
and all the elements $a_{ii}$ are arbitrary real numbers.

{\bf 2.} For a connection matrix $J^{(0)}$ with zero diagonal elements, the set of 
the fixed points coincides with the set of the local minima of the energy 
functional
$$FP(J^{(0)})=LM(J^{(0)}). \eqno(7)$$

{\bf 3.} Let all the elements $a_{ii}$ of the diagonal matrix (6) be positive, then
$$FP(J^{(0)}+A)\supseteq FP(J^{(0)})\supseteq FP(J^{(0)}-A). \eqno(8)$$ 

{\sl The proof of Theorem 1:}  Let's write the energy of the state 
$\vec \sigma$, extracting the contribution of the $l$th spin. Up to the 
positive factor we obtain:
$$E(\vec \sigma) \propto -\sum_{i,j\ne l} J_{ij}\sigma_i\sigma_j + J_{ll}
 -2\sigma_l h_l. \eqno(9)$$
The state $\vec \sigma$  will be the local minimum of the energy functional 
if and only if the system of the inequalities (10) is fulfilled for all 
states $\vec \sigma^{(l)}$ from the vicinity of the state $\vec \sigma$:
$$E(\vec \sigma^{(l)})-E(\vec \sigma) \propto \sigma_lh_l -J_{ll} =
\sigma_l\sum_{j\ne l}J_{lj}\sigma_j \ge 0,\quad l=1,2,\ldots,n. \eqno(10)$$
It is evident that the inequalities (10) do not depend on the values of the 
diagonal elements. The conditions of their fulfillment are defined by the 
off-diagonal part of the matrix $J$. By this the first item of the Theorem is 
proved. Moreover, it follows from the inequalities (10) that for nonnegative   
$J_{ll}$ any local minimum of the energy functional is also a fixed point 
of the network:
$$LM(J)\subseteq FP(J) \mbox{ when } J_{ll} \ge 0,\quad l=1,2,\ldots,n. 
\eqno(11)$$

Then, let a state $\vec \sigma^*$ be a fixed point of the network: 
$\sigma_l^*h_l \ge 0,\quad l=1,2,\ldots,n$. With the help of Eq.(9) we obtain:
$$E(\vec \sigma^{(l)})-E(\vec \sigma^*)\propto \sigma_l^*h_l-J_{ll}\left\{
\begin{array}{rccl}\ge&0,&\mbox{if}&J_{ll}=0\\
?&0,&\mbox{if}&J_{ll}>0\\ >&0,&\mbox{if}&J_{ll}<0\end{array}\right.$$
In other words, for nonpositive $J_{ll}$ any fixed point of the network is a 
local minimum of the energy functional:
$$LM(J)\supseteq FP(J) \mbox{ when } J_{ll}\le0,\quad l=1,2,\ldots,n.\eqno(12)$$
Combined with the proved first item of the Theorem, Eqs. (11) and (12) 
justify the correctness of Eqs. (7) and (8). Thus, the proof of the Theorem 
is finished.

In fact, to some extent the Theorem 1 permits regulating the set of the fixed 
points of the network. Let's explain this statement. If the matrix $J^{(0)}$  
is transformed by the diagonal matrix $A,\quad J(A)=J^{(0)}+A$, then in 
accordance with the Theorem 1, the set of the local minima of the energy 
functional is not changed. But this transformation affects the set of the 
fixed points of the $J(A)$-network. Indeed, if all the 
matrix elements $a_{ii}$ are positive, the set of the $J(A)$-network fixed 
points extends as compared with $FP(J^{(0)})$. The last is true due to the 
appearance of the new fixed points which are not the local minima. If, on the 
contrary, all the matrix elements $a_{ii}$ are negative, the set of the fixed 
points of the  $J(A)$-network narrows as compared with $FP(J^{(0)})$: some 
states, remaining, as they were, the local minima of the energy functional, 
cease to be the fixed points.

The last statement allows to suggest a simple method for the elimination of 
the unnecessary fixed points of the network. Let's formulate it in the form 
of a theorem.
\vskip 5mm
{\bf Theorem 2}

Let the fixed points of a $J^{(0)}$-network be numbered in such a way that
$$E(\vec \sigma^{(1)}\le E(\vec \sigma^{(2)}\le \ldots \le E(\vec \sigma^{(k)} 
<E(\vec \sigma^{(k+1)}\le \ldots $$
(to simplify the writing, here we omit the superscript "*" in the notations of
the fixed points). Let $A$ be a diagonal matrix whose elements are defined by 
the equalities

$$a_{ii}=\min_{l=1,k} \sigma_i^{(l)}h_i(\vec \sigma^{(l)}),
\quad i=1,2,\ldots n,\eqno (13)$$
where $h_i(\vec \sigma)$ is the potential (2) which acts on the $i$th spin in 
the $J^{(0)}$-network. Then 
$$FP(J^{(0)}-A)=\{\vec \sigma^{(1)},\vec \sigma^{(2)},\ldots,\vec \sigma^{(k)}\}.$$

{\sl The proof of Theorem 2:} Since all $a_{ii}$ from Eq. (13) are positive, 
the set of the $(J^{(0)}-A)$-network fixed points, due to Theorem 1, can be 
only narrower in comparison with the set $J^{(0)}$-network fixed points. And 
from the definition (4) of a fixed point, it follows that the state 
$\vec \sigma^{(l)}$ will be a $(J^{(0)}-A)$-network fixed point if and 
only if the system of the inequalities
$$\sigma_i^{(l)}h_i(\vec \sigma^{(l)})-a_{ii} \ge 0,\quad i=1,2,\ldots,n \eqno(14)$$
is fulfilled. When the definition (13) is taken into account, it is evident 
that for any fixed point $\vec \sigma^{(l)}$ with the $l\le k$ the system of 
the inequalities (14) is fulfilled. Consequently, the first $k$ states of 
the $\vec \sigma^{(l)}$ are the $(J^{(0)}-A)$-network fixed points. On 
the other hand, by proceeding from Eq.(5) for the energy and taking into 
account that the $J^{(0)}$-network fixed points are strictly ordered with
respect to the energy increase, it is easy to see that at least for one of 
the coordinates of the state $\vec \sigma^{(l)}$ with $l>k$, the 
inequalities (14) are not fulfilled. Consequently, the states $\vec \sigma^{(l)}$
with $l>k$ are not the $(J^{(0)}-A)$-network fixed points.

{\bf Remark.}  From Eq.(9) it can be easily shown that even under the 
sequential dynamics the evolution of a network with a connection matrix whose 
diagonal elements are negative can be accompanied by the energy increase. 
As a result, even under the sequential dynamics limit cycles can be formed 
for such a network! From this point of view, the connection matrices with 
negative diagonal elements are absolutely nonphysical. But Theorem 2 gives 
a simple and effective method to eliminate high energy fixed points. 
In some cases this method can be very useful. In particular, it is 
well-known~\cite{Pers,Lit2,Kant}, that for a network with the projection
connection matrix the energies of the spurious fixed points are larger then
the energies of the memorized patterns. Consequently all such spurious fixed
points can be easily eliminated with the help of Theorem 2.

\section{Projection connection matrix}

{\bf 1).} Let 
$$\vec \xi^{(l)}=(\xi^{(l)}_1,\xi^{(l)}_2,\ldots,\xi^{(l)}_n),\quad 
l=1,2,\ldots,p,\eqno(15)$$
be $p$ preassigned configuration vectors which we would like to have as a 
network fixed points (such vectors are usually called the {\sl memorized patterns}). 
It is known~\cite{Pers}, that it can be easily done if a matrix $P$ of orthogonal 
projection onto the linear subspace $\Lambda_{<\vec \xi^{(1)},\vec \xi^{(2)},
\ldots,\vec \xi^{(p)}>}$, spanned by the $p$ memorized patterns $\vec \xi^{(l)}$,
is taken as a connection matrix. The matrix $P$ is symmetric and nilpotent one:
$P^T=P,\quad P^2=P$. 
Besides, by definition $P\vec \xi^{(l)}=\vec \xi^{(l)},\quad l=1,2,\ldots,p$. 
Consequently, the vectors $\vec \xi^{(l)}$ are not only the fixed points of 
the $P$-network, but provide the global minimum of the energy functional (5):
$$E(\vec \xi^{(l)})=-\frac{(P\vec \xi^{(l)},\vec \xi^{(l)})}n=-1,\quad l=1,2,\ldots,p.$$
But, it is known from experience, that, as a rule, the $P$-network has 
additional fixed points which are called {\sl spurious} fixed points. Their 
number is much larger that for the network with Hebb's connection matrix. 
And the worst is that not all the $P$-network fixed points are the local 
minima of the energy functional~\cite{Kant}. Theorem 1 helps to clarify the situation.

{\bf 2).} For simplicity we assume that $p$ memorized patterns $\vec \xi^{(l)}$  
are linearly independent vectors. We introduce a rectangular $(p\times n)$-
matrix $\Xi$ whose rows are memorized patterns $\vec \xi^{(l)}$:
$$\Xi=\left(\begin{array}{cccc}
\xi_1^{(1)}&\xi_2^{(1)}&\ldots&\xi_n^{(1)}\\\xi_1^{(2)}&\xi_2^{(2)}&\ldots&\xi_n^{(2)}\\
\ldots&\ldots&\ldots&\ldots\\\xi_1^{(p)}&\xi_2^{(p)}&\ldots&\xi_n^{(p)}\end{array}
\right) \eqno(16)$$
Then the matrix of the orthogonal projection onto the linear subspace 
$\Lambda_{<\vec \xi^{(1)},\vec \xi^{(2)},\ldots,\vec \xi^{(p)}>}$ is 
$$P=Y\Xi, \eqno(17)$$
where $Y$ is the $(n\times p)$-matrix that is pseudoinverse of the matrix 
$\Xi$. Apropos of the construction of pseudoinverse matrices, see~\cite{Pers,Lit2}. 
We only want to mention, that the columns of the matrix $Y$ are $n$-dimensional 
vectors $\vec y^{(l)}$ such that $(\vec y^{(l)},\vec \xi^{(l')})=\delta_{ll'}$,
where $\delta_{ll'}$ is the Kronecker symbol; $\vec y^{(l)}$ are also the 
linearly independent vectors.

The diagonal elements of the matrix $P$ are positive. Indeed, they are equal 
to the squares of the projections onto the subspace    
$\Lambda_{<\vec \xi^{(1)},\vec \xi^{(2)},\ldots,\vec \xi^{(p)}>}$
of the $n$-dimensional Cartesian unit vectors 
$\vec e^{(i)}=(0,\ldots,1,\ldots,0)$:
$$ P_{ii}=(P\vec e^{(i)},\vec e^{(i)})=\parallel P\vec 
e^{(i)}\parallel^2=d_i^2,\quad i=1,2,\ldots,n.\eqno(18) $$
The values $d^2_i$ must be nonzero, otherwise the vectors 
$P\vec e^{(i)}=\sum_{j=1}^p\xi_i^{(j)}\vec y^{(j)}$, which are the sums of the 
linearly independent vectors, would be zero. Then, according to Theorem 1, 
the set of the $P$-network fixed points will really be wider than the set of 
the local minima of the energy functional.

If instead the $P$-network we consider the $P^{(0)}$-network,
$$P^{(0)}=P-D,\mbox{ where } D=diag\{d_1^2,d_2^2,\ldots,d_n^2\}\eqno(19)$$
only the local minima of the energy functional will be the fixed points of the 
$P^{(0)}$-network. Since the memorized patterns provide the global minimum of 
the energy functional, they will necessarily be the $P^{(0)}$-network fixed 
points. In addition, the $P^{(0)}$-network has no nonphysical fixed points, 
i.e. those which are not the local minima of the energy functional.

Let's show that the set of the $P^{(0)}$-network fixed points has a structure 
which resembles the structure of the set of the fixed points for Hopfield's 
model~\cite{Lit2}.
\vskip 5mm
{\bf Theorem 3}

All $n$ neurons can be enumerated in such a way that all the $P^{(0)}$-network 
fixed points will be of a "piecewise constant" kind:
$$\vec \sigma^* =(\underbrace{\varepsilon_1,\varepsilon_1,\ldots,
\varepsilon_1}_{n_1},\underbrace{\varepsilon_2,\varepsilon_2,\ldots,
\varepsilon_2}_{n_2},\ldots,\underbrace{\varepsilon_v,\varepsilon_v,\ldots,
\varepsilon_v}_{n_v}), \eqno(20)$$
where $\varepsilon_i=\{\pm 1\}$. The number $v$, the composition and the 
dimension $n_i$ of constant sign intervals are defined uniquely by the 
memorized patterns matrix $\Xi$ (16).

{\sl The proof of Theorem 3}. For simplicity we assume that the number of the 
neurons $n$ is much larger than the number of the memorized patterns $p$. For 
definiteness, we assume that $n>2^p$ . Then it is evident that not all $n$ 
$p$-dimensional 
column-vectors
$$\vec \xi_i=\left(\begin{array}{c}\xi_i^{(1)}\\\xi_i^{(2)}\\\vdots\\
\xi_i^{(p)}\end{array}\right),\quad i=1,2,\ldots,n\eqno(21)$$
of the matrix$\Xi$ will be different. Some of them will be repeated
\footnote{The $p$-dimentional column-vectors $\vec \xi_i$  (21) are labelled 
with subscripts in contrast to the $n$-dimensional memorized patterns 
$\vec \xi^{(l)}$ (15), which are labelled with superscripts.}. Let's assume 
that $n$ column-vectors $\vec \xi_i$ break up into $v$ groups. Each group 
consists of some identical vectors: $n_1$ identical vectors belong to the 
first group; $n_2$ identical vectors belong to the second group and so on. 
And $n_1+n_2+\ldots+n_v=n$. Without loss of generality it may be assumed that 
the neurons of the network can be numbered in such a way that the 
matrix $\Xi$ will have a form:
$$\Xi=\left(\begin{array}{cccccccccc}
\xi_1^{(1)}&\ldots&\xi_1^{(1)}&\xi_2^{(1)}&\ldots&\xi_2^{(1)}&\ldots&\xi_v^{(1)}&\ldots&\xi_v^{(1)}\\
\xi_1^{(2)}&\ldots&\xi_1^{(2)}&\xi_2^{(2)}&\ldots&\xi_2^{(2)}&\ldots&\xi_v^{(2)}&\ldots&\xi_v^{(2)}\\
\cdots&\cdots&\cdots&\cdots&\cdots&\cdots&\cdots&\cdots&\cdots&\cdots\\
\xi_1^{(p)}&\ldots&\xi_1^{(p)}&\xi_2^{(p)}&\ldots&\xi_2^{(p)}&\ldots&\xi_v^{(p)}&\ldots&\xi_v^{(p)}
\end{array}\right)\eqno(22)$$
It was shown in~\cite{Lit2} that when the projection matrix $P$ acts onto the 
configuration vector $\vec \sigma$ the result is
$$P\vec \sigma=\sum_{l=1}^pm^{(l)}(\vec \sigma)\vec \xi^{(l)},$$
where $m^{(l)}(\vec \sigma)$ depends only on the state $\vec \sigma$.  
If $\vec \sigma^*$ is a fixed point of the $P^{(0)}$-network, 
its coordinates $\sigma^*$ must satisfy the system of equations:
$$sign(P\vec \sigma^*-D\vec \sigma^*)\equiv sign\left(\sum_{l=1}^pm^{(l)}(
\vec \sigma^*)\xi_i^{(l)}-d_i^2\sigma_i^*\right)=\sigma_i^*,\quad i=1,2,\ldots,n.
\eqno(23)$$
With respect to the form of the matrix $\Xi$ (see Eq.(22) ), it is easy to 
understand that, for example, for the first $n_1$ values of the subscript $i$ 
all the sums $\sum_{l=1}^pm^{(l)}(\vec \sigma^*)\xi_i^{(l)}$ are the same:
$\sum_{l=1}^pm^{(l)}(\vec \sigma^*)\xi_i^{(l)}=M,\quad i=1,2,\ldots,n_1$. 
From here it immediately follows, that the first $n_1$ coordinates of the 
fixed point $\vec \sigma^*$ must be equal. Indeed, let's contrarily assume 
that, for example, $\sigma^*_1=1 \mbox{ and }\sigma_2^*=-1$. Then from Eqs. 
(23) it follows that two inequalities 
$$\left\{\begin{array}{c}M\ge d_1^2\\-M\ge d_2^2\end{array}\right.$$
have to be fulfilled simultaneously. But this is impossible.

Consequently, the first $n_1$ coordinates of the $P^{(0)}$-network fixed 
point are equal to one another. Similarly, it can be proved that the next $n_2$
coordinates of the fixed point are equal to one another too, and so on up to 
the equality among the last $n_v$ coordinates. The Theorem is proved.

{\bf 3).} Thus, the $P^{(0)}$-network possesses some attractive 
properties. Namely, all the memorized patterns will necessarily be its fixed 
points. Moreover, all the fixed points are, firstly, the local minima of the 
energy functional; secondly, they are of "piecewise constant" kind given by 
Eq. (20). It is known~\cite{Lit2,Ezh} that the last two of the above mentioned 
properties are also typical for the fixed points of the Hopfield model, that 
is, for a network with the Hebb connection matrix. In our following discussions 
we will have in mind either the $P^{(0)}$-network, or the Hopfield model.

While Eq.(20) is only the necessary condition for a configuration to be a 
fixed point of a network, it restricts sufficiently the circle of the 
"applicants". Let's provide some estimates.

If $v$ is the number of {\sl different} column-vectors of the matrix $\Xi$ and $q$ 
is the number of the fixed points of the related network, it is clear that 
$q\le 2^v$. Next, when the number of the memorized patterns $p$ is given, 
a natural estimate for $v$ is: $v\le 2^p$. But, as it was shown in~\cite{Lit2}, 
actually we have:
$$q\le 2^{2^{p-1}}.\eqno(24)$$

Eq. (24) restricts the number of the fixed points from above. It is easy to 
verify that when $p=2$, the inequality (24) transforms into the equality, 
and $q$ just equals 4. But even for $p=3$  the right-hand side of Eq.(24) 
is 16, and in~\cite{Lit2} it was shown that for $p=3$  the maximum possible 
number of the network fixed points is 14. This result coincides with the 
one of~\cite{Bal1}. Other estimates for the fixed points number can be found 
in~\cite{Bal2,Bal3,Kuhl} .

Partially this work was supported by RFBR grant No. 95-01-01191.

\end{document}